\documentclass[twocolumn,twoside]{revtex4-1}
\usepackage{graphicx}
\usepackage{fancyhdr}
\def\fermi{{\it Fermi }}

\begin{document}

\title{A Southern Sky Survey with Fermi LAT and ASKAP}

\author{Robert A. Cameron}
\affiliation{SLAC/KIPAC, Stanford, CA 94025, USA}

\author{on behalf of the \fermi Large Area Telescope Collaboration}
\author{on behalf of the ASKAP VAST Science Survey Team}
\affiliation{ }

\begin{abstract}
We present the prospects for a future joint gamma-ray and radio survey 
of southern hemisphere sources using the \fermi Large Area Telescope 
(LAT) and the upcoming Australian Square Kilometre Array Pathfinder 
(ASKAP) radio telescope.  ASKAP is a next generation radio telescope 
designed to perform surveys at GHz frequencies at a much higher survey 
speed than previous radio telescopes, and is scheduled to start 
engineering observations in 2011.  The survey capabilities of both \fermi 
LAT and ASKAP are described, and the planned science surveys for 
ASKAP are summarized. 
We give some expected details of the Variable and Slow Transient (VAST) 
survey using ASKAP, which will search for transients on timescales from 
5 seconds to years. Some observational properties of faint and transient 
sources seen at gamma-ray and radio wavelengths are summarized, 
and prospects and strategies for using ASKAP survey data for LAT source 
counterpart identification are summarized. 
\end{abstract}

\maketitle

\thispagestyle{fancy}

\section{The Large Area Telescope}
\begin{figure}
\includegraphics[width=82mm]{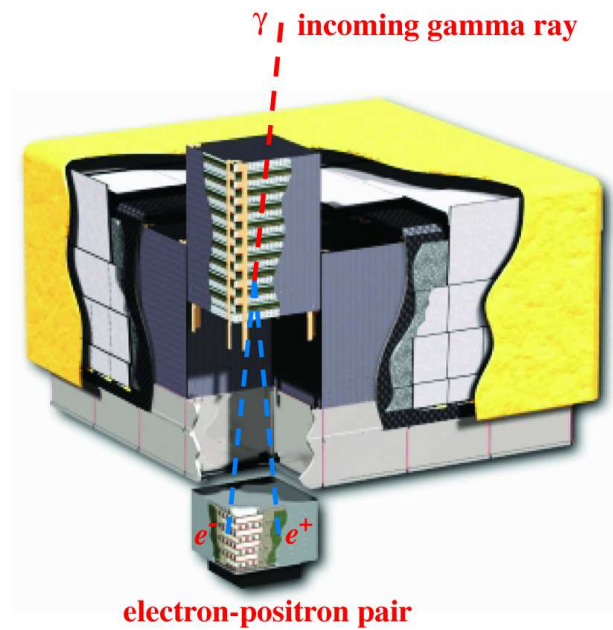}
\caption{Cutaway diagram of the LAT showing a Tracker, Calorimeter and ACD.
The LAT dimensions are approximately 1.8m x 1.8m x 0.75m.}
\label{fig:LAT}
\end{figure}

The Large Area Telescope (LAT, Figure~\ref{fig:LAT}, \cite{LATpaper}) 
on \fermi is a pair-conversion telescope, 
having a 4x4 grid of detectors (TKR+CAL) surrounded by an ACD. TKRs 
measure photon direction, CALs measure photon energy. The ACD vetoes
charged particle background events. The major elements of the LAT are:
\begin{itemize}
\item Tracker (TKR): Stacked Si-strip detector layers with superb position
resolution and efficiency, with low-power readout. Multiple tungsten foils
allow good angular resolution while providing high conversion efficiency.
1.5 X$_0$ total depth.
\item Calorimeter (CAL): Hodoscopic array of CsI crystals with PIN-diode
readout. Segmentation provides shower imaging for improved energy 
reconstruction and background rejection. Radiation length: 8.4 X$_0$ total 
at normal incidence.
\item Anti-Coincidence Detector (ACD): Plastic scintillator array for high
efficiency charged-particle detection and minimal self-veto of gamma-rays.
\item Electronics, DAQ,Trigger: Process and filter events from the LAT. Perform
on-board searches for gamma-ray bursts.
\end{itemize}
The main performance specifications of the LAT are summarized in 
Table~\ref{tab:LATperf}.

\begin{table}
\begin{center}
\caption{LAT Performance Summary}
\begin{tabular}{|l|c|}
\hline ~Energy Range (GeV) & 0.02 to $>300$ \\
\hline ~Effective Area (on-axis, $>1$ GeV)& 8000 cm$^2$ \\
\hline ~Angular Resolution (@ 10 GeV)~ & 0.1$^\circ$ \\
\hline ~Energy Resolution (on-axis, $> 0.1$ GeV)~ & $<15$\% \\
\hline ~Event Absolute Timing & $<10 \mu$s \\
\hline ~Field of View & 2.4 sr \\
\hline ~Deadtime per Event & $ 27 \mu$s \\
\hline
\end{tabular}
\label{tab:LATperf}
\end{center}
\end{table}

\section{The LAT Sky Survey}

\begin{figure}
\includegraphics[width=82mm]{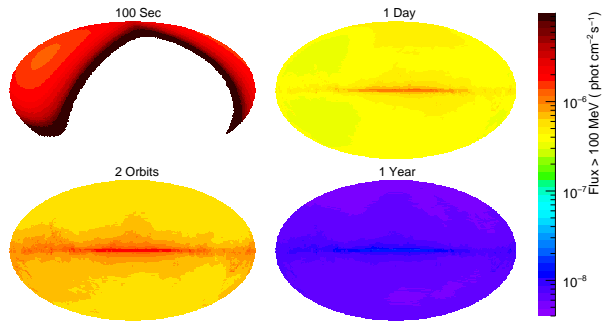}
\caption{Sky Coverage and Sensitivity of the LAT Sky Survey on timescales 
of 100 seconds, 2 orbits, 1 day and 1 year.}
\label{fig:LATskysurvey}
\end{figure}

The field of view of the LAT covers $\sim$20\% of the sky at any instant, and up 
to 75\% of the sky every orbit. In scanning mode the entire sky is observed 
every 2 orbits ($\sim$3 hours).  Figure~\ref{fig:LATskysurvey} shows the LAT 
sensitivity to point sources across the sky (in galactic coordinates) for 
integration periods of 100 seconds, 2 orbits ($\sim$192 minutes), 1 day and 
1 year. The LAT has less sensitivity on the galactic plane because of the 
higher diffuse emission there.

\begin{figure}
\includegraphics[width=82mm]{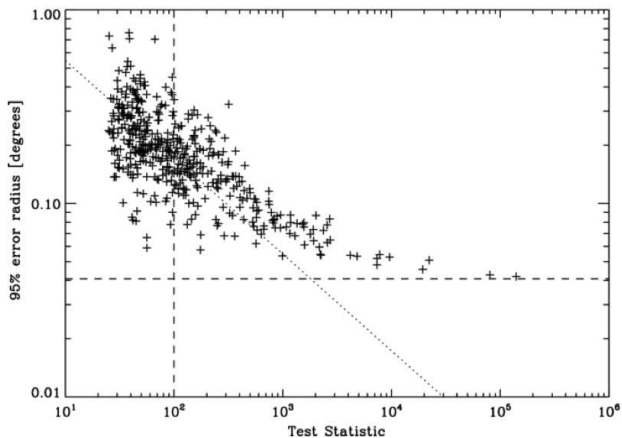}
\caption{LAT source position uncertainty, the 95\% confidence radius on 
measured source positions, as a function of the LAT source measurement 
Test Statistic, for bright sources detected in the first 3 months of the LAT sky 
survey \cite{brightsourcelist}.}
\label{fig:LAT95Rad}
\end{figure}

\begin{figure}
\includegraphics[width=82mm]{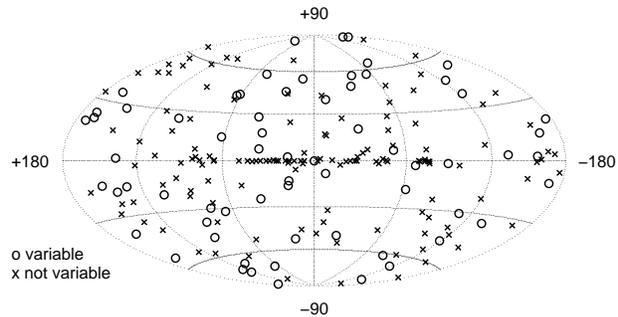}
\caption{The positions in Galactic coordinates of the bright sources detected
in the first 3 months of the LAT sky survey \cite{brightsourcelist}.}
\label{fig:LATskymap}
\end{figure}

The source localization capability of the LAT is shown in Figure~\ref{fig:LAT95Rad}, 
which plots the 95\% confidence radius for source positions as a function of 
the source Test Statistic (measurement significance), from the bright source 
list measured in the first three months of LAT survey data \cite{brightsourcelist}. The brightest 
sources have a position uncertainty near 0.04¡, limited by systematic 
uncertainties, while fainter sources can have a position uncertainty of tens 
of arc-minutes. Figure~\ref{fig:LATskymap} shows the distribution of sources 
in the 3-month bright source list, showing variable and non-variable sources. 
Measuring correlated variability at other wavelengths is one approach to 
counterpart identification of LAT sources.

\section{ASKAP}

Australia and South Africa are the two candidate sites for the future 
Square Kilometre Array (SKA) radio telescope, the next generation 
successor to the current largest radio interferometer array telescopes. 
Final site selection for the SKA is planned for 2012, and demonstrator 
radio telescopes are currently being constructed at each candidate site. 

\begin{figure}
\includegraphics[width=78mm]{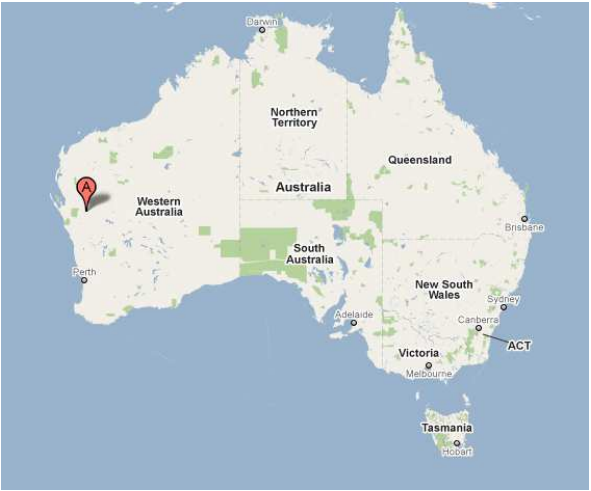}
\caption{The site of the ASKAP telescope array center, in Western Australia. 
The array will be centered at Latitude = 26.7$^\circ$ S, Longitude = 
115.5$^\circ$ E.}
\label{fig:ASKAPsite}
\end{figure}

The ASKAP (Australian Square Kilometre Array Pathfinder) radio 
telescope is being developed under the guidance of the ASKAP 
International Collaboration, led by the Australia Telescope National 
Facility (ATNF) in Australia \cite{ASKAP}. The member institutions of 
the ASKAP International Collaboration are
\renewcommand{\labelitemii}{$\star$}
\begin{itemize}
\item Commonwealth Scientific and Industrial Research Organization 
(CSIRO), Australia
\begin{itemize}
   \item Australia Telescope National Facility (ATNF)
   \item Information and Communication Technologies Centre
\end{itemize}
\item University of Western Australia
\item Herzberg Institute of Astrophysics (Canada)
\item ASTRON (Netherlands)
\item Max Planck Institute (Germany)
\item Auckland University of Technology (New Zealand)
\end{itemize}
ASKAP will be located at the candidate site for the SKA central array in 
Murchison, Western Australia, (Figure~\ref{fig:ASKAPsite}) and will be 
co-located with the Murchison Wide Field Array (MWA), a 
low-frequency (80-300 MHz) dipole array. The 
first ASKAP antenna will be on site in late 2009. ASKAP will start 
partial-array observations in 2011, and full-array observations are
scheduled to start in late 2012. At least 75\% of ASKAP observing time 
will be used for several large Survey Science Projects during the first 5 
years of operation: 2012 Ð-- 2017.

\section{ASKAP Array Configuration}

\begin{figure}
\includegraphics[width=77mm]{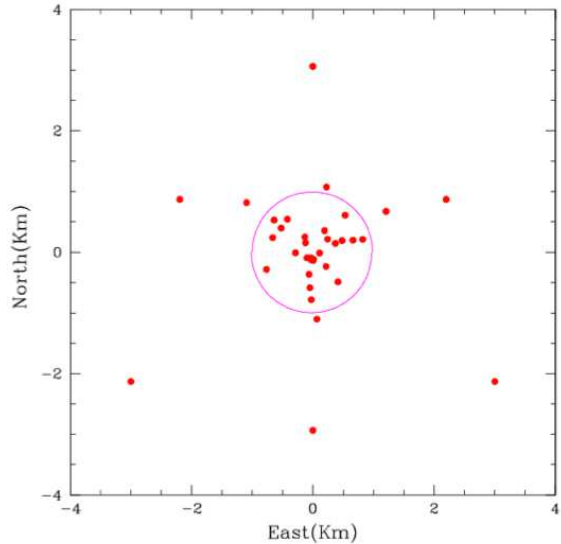}
\caption{The ASKAP telescope array layout.}
\label{fig:ASKAPlayout}
\end{figure}

\begin{figure}
\includegraphics[width=75mm]{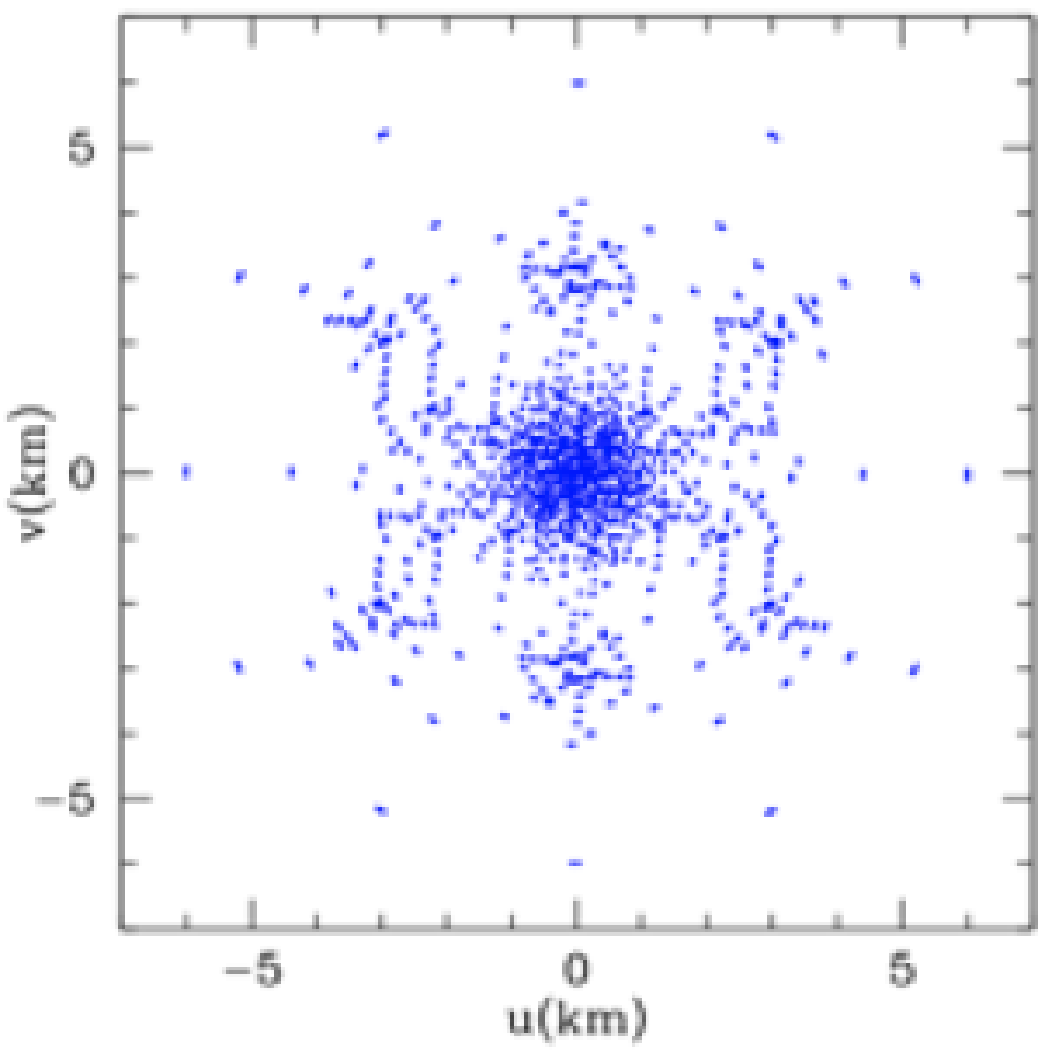}
\caption{The {\it uv} visibility distribution from the ASKAP telescope array 
for a 12 minute observation of a source at declination $-30^\circ$.}
\label{fig:uv12m}
\end{figure}

\begin{figure}
\includegraphics[width=75mm]{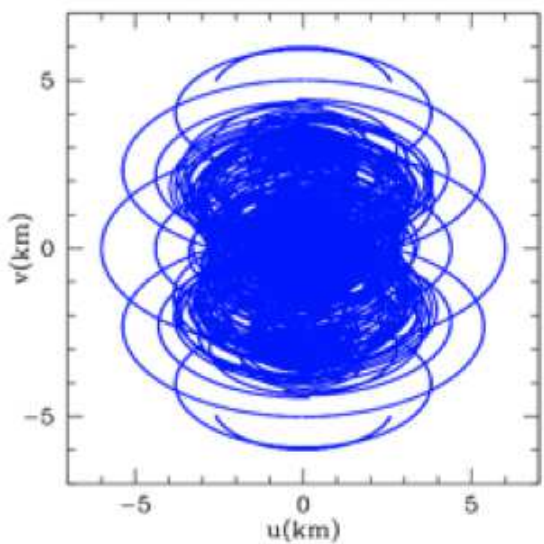}
\caption{The {\it uv} visibility distribution from the ASKAP telescope array 
for a 10 hour observation of a source at declination $-30^\circ$.}
\label{fig:uv10h}
\end{figure}

The key performance parameters of the ASKAP telescope array are given in 
Table~\ref{tab:ASKAPconfig}. The layout of the ASKAP dishes in the array is 
shown in Figure~\ref{fig:ASKAPlayout}, and examples of the {\it uv} visibility 
coverage of the array for a source at declination $-30^\circ$ are shown in 
Figure~\ref{fig:uv12m} for a 12 minute observation, and in 
Figure~\ref{fig:uv10h} for a 10 hour observation.

\begin{table}
\begin{center}
\caption{ASKAP Performance Summary \cite{ASKAP}}
\begin{tabular}{|l|c|}
\hline ~Number of dishes & 36 \\
\hline ~Dish Diameter (m) & 12 \\
\hline ~Dish Area (m$^2$) & 113 \\
\hline ~Total Collecting Area (m$^2$)~ & 4072 \\
\hline ~Aperture Efficiency & 0.8 \\
\hline ~System Temperature & 50 \\
\hline ~Field of View (deg$^2$) & 30 \\
\hline ~Frequency Range (MHz) & ~700 - 1800~ \\
\hline ~Bandwidth (MHz) & 300 \\
\hline ~Maximum Channels & 16384 \\
\hline ~Maximum Baseline (km) & 6 \\
\hline
\end{tabular}
\label{tab:ASKAPconfig}
\end{center}
\end{table}

%\section{ASKAP Survey Capabilities}
By using wide-field-of-view phased array feeds to produce a large 
instantaneous field of view, ASKAP will be able to survey the whole 
sky vastly faster than is possible with the current generation of radio 
telescopes. Table~\ref{tab:ASKAPsurvspd} summarizes the capabilities 
of the ASKAP array in sky survey speed for continuum, spectral line 
and surface brightness measurements, as a function of angular 
resolution. Similarly, Table~\ref{tab:ASKAPsurvsens} summarizes 
the 1$\sigma$ sensitivity for 1 hour integrations in continuum, 
spectral line and surface brightness measurements.

\begin{table*}
\begin{center}
\caption{ASKAP Survey Speed Capabilities (deg$^2$/hour)}
\begin{tabular}{|l|c|c|c|c|c|c|}
\hline  ~{\bf Angular Resolution (arcsec)} & {\bf 10} & {\bf 18} & {\bf 30} & {\bf 90} & ~{\bf 180}~ \\
\hline  ~Continuum Survey Speed (300 MHz, 100 $\mu$Jy) & ~220~ & ~361~ & ~267~ & ~54~ & 17 \\
\hline  ~Line survey Speed (100 kHz, 5 mJy) & 184 & 301 & 223 & 45 & 14 \\
\hline  ~Surface Brightness Survey Speed (5 kHz, 1$^\circ$K)~ & -- & -- & 1.1 & 18 & 94 \\
\hline
\end{tabular}
\label{tab:ASKAPsurvspd}
\end{center}
\end{table*}

\begin{table*}
\begin{center}
\caption{ASKAP Survey Sensitivity (1 hour observation, 1$\sigma$)}
\begin{tabular}{|l|c|c|c|c|c|c|}
\hline  ~{\bf Angular Resolution (arcsec)} & {\bf 10} & {\bf 18} & {\bf 30} & {\bf 90} & {\bf 180} \\
\hline  ~Continuum Survey Sensitivity (300 MHz) ($\mu$Jy/beam)~ & 37 & 29 & 34 & 74 & 132 \\
\hline  ~Line survey Sensitivity (100 kHz) (mJy/beam) & ~2.1~ & ~1.6~ & ~1.9~ & ~4.1~ & 7.3 \\
\hline  ~Surface Brightness Survey Sensitivity (5 kHz) ($^\circ$K)~ & 51 & 12 & 5.2 & 1.3 & ~0.56~ \\
\hline
\end{tabular}
\label{tab:ASKAPsurvsens}
\end{center}
\end{table*}

\section{The Variable and Slow Transient Survey}

Following a request for science survey proposals starting in 2008, 10 proposals 
were selected in September 2009 to be performed in the first 5 years of ASKAP operation:
\begin{itemize}
\item EMU: Evolutionary Map of the Universe (PI: Norris)
\item WALLABY: Widefield ASKAP L-Band Legacy All-Sky Blind Survey (PIs: Koribalski and Staveley-Smith)  
\item ASKAP-FLASH: The First Large Absorption Survey in H I (PI: Sadler)
\item VAST: An ASKAP Survey for Variables and Slow Transients  (PIs: Murphy and Chatterjee
\item GASKAP: The Galactic ASKAP Spectral Line Survey (PI: Dickey)
\item POSSUM: Polarization Sky Survey of the Universe's Magnetism (PIs: Gaensler, Taylor and Landecker)
\item CRAFT: The Commensal Real-time ASKAP Fast Transients survey (PIs: Dodson and Macquart
\item DINGO: Deep Investigations of Neutral Gas Origins (PI: Meyer)
\item The High Resolution Components of ASKAP: Meeting the Long Baseline Specifications for the SKA (PI: Tingay)
\item COAST: Compact Objects with ASKAP:  Surveys and Timing (PI: Stairs) 
\end{itemize}
The Variable and Slow Transient (VAST) survey is particularly relevant to identification 
of transient sources detected with Fermi LAT, since it concentrates on the detection and
measurement of variable and transient radio sources on timescales down to 5 seconds. 
The VAST proposal included 3 sub-surveys:
\begin{itemize}
\item VAST-Wide: 10,000 square degrees observed every day for 
2 years (rms sensitivity = 0.5 mJy/beam). 
\item VAST-Deep: 10,000 square degrees (rms sensitivity = 0.05 mJy/beam). 
Fields revisited 8 times at irregularly spaced intervals. 
\item VAST-GP: 750 square degrees along the Galactic plane plus LMC and 
SMC (rms sensitivity = 0.1 mJy/beam) repeated weekly for 1 year.
\end{itemize}
The VAST survey will be complementary to other ASKAP surveys: e.g. the EMU survey 
has a large single epoch deep survey (rms sensitivity 10 $\mu$Jy/beam). The VAST 
proposal also included commensal transient searches in other ASKAP survey data. 
Since the 10 accepted surveys exceed the 75\% goal for ASKAP dedicated survey 
time, some descope or merging of the VAST surveys may be negotiated. 
The VAST daily radio survey of much of the southern hemisphere sky will greatly
benefit the identification and study of faint and/or variable sources detected by the 
Fermi LAT. The Fermi mission will end its first 5 years of operation in 2013, with 
the expected additional five years for the 10-year Fermi mission overlapping well 
with the ASKAP surveys.

\section{Radio and Gamma-ray Source Properties}
\begin{figure}
\includegraphics[width=82mm]{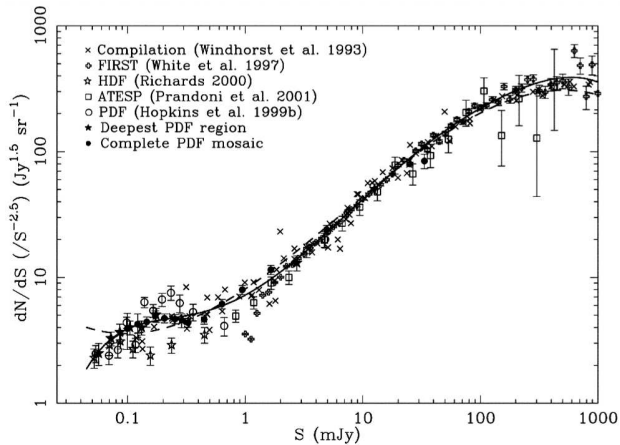}
\caption{1.4 GHz differential radio source counts, from multiple catalogs (from \cite{radiosourcecounts}).}
\label{fig:radiosourcecounts}
\end{figure}

The bright source list from the first 3 months of the Fermi LAT sky survey contained 
205 sources and included 66 sources showing significant variability on weekly or 
longer timescales. 37 of the 205 bright sources (18\%) had no known obvious 
counterpart at other wavelengths, with most of the unassociated sources being 
on or near the galactic plane. As the Fermi survey continues and fainter sources
are detected, the problem of source identification will worsen: approximately
50\% of the sources in the preliminary LAT first year source catalog have no 
association at other wavelengths (preliminary result). Deep, large area radio
surveys can provide many additional source associations. However, current 
deep surveys (e.g. \cite{radiosourcecounts}, Figure~\ref{fig:radiosourcecounts})
show that the sky surface density of 1mJy or brighter sources at 1.4 GHz will 
potentially result in several candidate counterparts per faint LAT source.

\begin{figure}
\includegraphics[width=73mm]{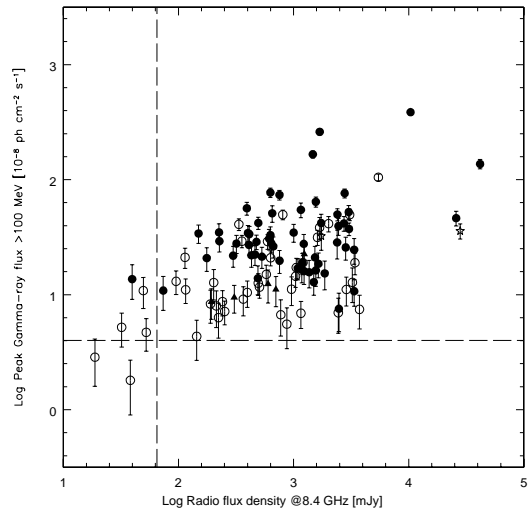}
\caption{Gamma-ray to radio source flux correlation for Flat Spectrum Radio Quasars 
(filled circles) and BL Lacs (open circles), for the LAT bright AGN source list 
\cite{LATbrightAGNsurvey}.}
\label{fig:gammaradiofluxcorr}
\end{figure}

Simple correlations of gamma-ray to radio fluxes or luminosities will not 
provide adequate counterpart identifications. Figure~\ref{fig:gammaradiofluxcorr} 
shows the scatter in the correlation of gamma-ray flux to radio flux for sources
in the LAT bright AGN source list \cite{brightsourcelist}. Additionally correlating 
source variability data from the LAT and  VAST surveys, for both galactic and 
extragalactic sources will aid radio counterpart identification to LAT sources, 
and will also aid in separating foreground radio variability effects such as 
scintillation from intrinsic source variations. We expect to apply both spatial 
and temporal analysis to LAT and ASKAP data for finding LAT source 
counterparts, and for their further study. Daily radio measurements from 
VAST should generally provide sufficient time sampling for correlation 
to gamma-ray source variability observed with the LAT.

\begin{acknowledgments}
This work is supported by Stanford University and the SLAC National Accelerator 
Laboratory under DOE contract DE-AC03-76SFO0515 and NASA grant NAS5-00147. 
Non-US sources of funding also support the efforts of \fermi LAT collaborators in 
France, Italy, Japan and Sweden.
\end{acknowledgments}


\begin{thebibliography}{9}   % Use for  1-9  references
%\begin{thebibliography}{99} % Use for 10-99 references

\bibitem{LATpaper}
Atwood, W. B. et al. 2009, ApJ, 697, 1071.

\bibitem{brightsourcelist}
Abdo, A. A. {\it et al.} 2009, ApJS, 183, 46.

\bibitem{ASKAP}
http://www.atnf.csiro.au/projects/askap/index.html

\bibitem{radiosourcecounts}
Hopkins A.M. {\it et al.} 2003, AJ, 125, 465.

\bibitem{LATbrightAGNsurvey}
Abdo, A. A. {\it et al.} 2009, ApJ, 700, 597.

\end{thebibliography}

\end{document}